\documentclass[prl, aps, twocolumn, superscriptaddress, groupedaddress, showpacs,a4paper]{revtex4} 

\usepackage{graphicx}  
\usepackage{bm}        
\usepackage{amssymb}   
\usepackage[latin1]{inputenc}
\usepackage[T1]{fontenc}
\usepackage[english]{babel}
\usepackage{subfigure}
\usepackage{epsfig}
\usepackage{verbatim}
\usepackage{setspace}
\usepackage{placeins}
\usepackage{amsmath}
\usepackage{mathrsfs}
\usepackage{slashed}
\usepackage[usenames, dvipsnames]{color}
\usepackage{graphics}
\usepackage{hyperref}
\usepackage[mediumspace,mediumqspace,grey,squaren]{SIunits}
\usepackage{amsfonts}
\hypersetup{
colorlinks=true,
linkcolor=blue,
linktoc=page,
citecolor=blue}

\begin{document}

\title{Annular Vortex Chain in a Resonantly Pumped Polariton Superfluid}
\author{T. Boulier}
\affiliation{Laboratoire Kastler Brossel, Universit\'{e} Pierre et Marie Curie, Ecole Normale Sup\'{e}rieure et CNRS,\\
UPMC Case 74, 4 place Jussieu, 75252 Paris Cedex 05, France}
\author{H. Ter\c{c}as}
\affiliation{Institut Pascal, PHOTON-N2, Clermont Universit\'e, Blaise Pascal University, CNRS,24 Avenue des Landais, 63177 Aubi\`ere Cedex, France}
\author{D. D. Solnyshkov}
\affiliation{Institut Pascal, PHOTON-N2, Clermont Universit\'e, Blaise Pascal University, CNRS,24 Avenue des Landais, 63177 Aubi\`ere Cedex, France}
\author{Q. Glorieux}
\affiliation{Laboratoire Kastler Brossel, Universit\'{e} Pierre et Marie Curie, Ecole Normale Sup\'{e}rieure et CNRS,\\
UPMC Case 74, 4 place Jussieu, 75252 Paris Cedex 05, France}
\author{E. Giacobino}
\affiliation{Laboratoire Kastler Brossel, Universit\'{e} Pierre et Marie Curie, Ecole Normale Sup\'{e}rieure et CNRS,\\
UPMC Case 74, 4 place Jussieu, 75252 Paris Cedex 05, France}
\author{G. Malpuech}
\email{malpuech@univ-bpclermont.fr}
\affiliation{Institut Pascal, PHOTON-N2, Clermont Universit\'e, Blaise Pascal University, CNRS,24 Avenue des Landais, 63177 Aubi\`ere Cedex, France}
\author{A. Bramati}
\email{bramati@lkb.umpc.fr}
\affiliation{Laboratoire Kastler Brossel, Universit\'{e} Pierre et Marie Curie, Ecole Normale Sup\'{e}rieure et CNRS,\\
UPMC Case 74, 4 place Jussieu, 75252 Paris Cedex 05, France}

\date{\today}

\begin{abstract}
We report the formation of a ring-shaped array of vortices after injection of angular momentum in a polariton superfluid. The angular momentum is injected by a $\ell=8$ Laguerre-Gauss beam, whereas the global rotation of the fluid is hindered by a narrow Gaussian beam placed at its center. In the linear regime a spiral interference pattern containing phase defects is visible. In the nonlinear (superfluid) regime, the interference disappears and the vortices nucleate as a consequence of the angular momentum quantization. The radial position of the vortices evolves freely in the region between the two pumps as a function of the density. Hydrodynamic instabilities resulting in the spontaneous nucleation of vortex-antivortex pairs when the system size is sufficiently large confirm that the vortices are not constrained by interference when nonlinearities dominate the system.
\end{abstract}

\maketitle

{\it Introduction.}$-$Quantized vortices have been extensively investigated in different fields of physics, such as superconductivity \cite{pismen}, matter-wave superfluids \cite{fetter} and nonlinear optics \cite{desyatnikov}. More recently, the discovery of polariton Bose-Einstein condensation in semiconductor microcavities \cite{kasprzak, balili, lai} has renewed the expectations of featuring superfluidity in quantum fluids of light \cite{carusotto}. As a consequence, the understanding of the mechanism of vortex nucleation and vortex dynamics in such systems becomes particularly important. Half-light, half-matter (i.e. polariton) fields are often described by nonlinear mean-field models, accounting for both Hamiltonian and dissipative terms. The phase symmetry inherent to these models is essential to the spontaneous formation of vortices and vortex lattices. Incoherent pump, in particular, does not destroy the global phase symmetry inherited from the Hamiltonian (conservative) part of the dynamics, which is associated to the conservation of the particle number. Therefore, vortices are natural topological solutions in interacting photonic systems \cite{weiss}. Indeed, the nucleation of vortices has been reported in several experiments \cite{lagoudakis, manni}. \par
Several schemes to produce vortices and vortex lattices have been proposed and realized. Liew et al \cite{liew} have proposed the formation of regular triangular lattices and the Penrose triangular lattices with coherently pumped polariton condensates, while Gorbarch and co-workers \cite{gorbach} have proposed to create robust half-vortex lattices in the optical-parametric-oscillator (OPO) scheme. Spontaneous self-ordered vortex-antivortex pairs have been recently reported in \cite{manni2} and Hivet et al \cite{PRB-Hivet} experimentally demonstrated the formation of regular vortex-antivortex lattices in square and triangle trapping potentials. Nevertheless, these schemes do not produce vortex lattices in the Abrikosov sense as they do not spontaneously form a group of self-ordered same-sign vortices. A closer attempt toward the production of such self-ordered, Abrikosov-like vortex lattices has been put forward in Ref. \cite{keeling}. In this ideal, disorder-free scheme, an important experimental limitation arises since the lattice rotates at very high speeds for which time-resolved interferometry measurements are quite difficult. In more realistic samples with disorder, vortices are pinned and the rotation can be stopped, but the vortex equilibrium positions are random as they depend on the disorder potential.  \par
To circumvent the previous issues, and to further improve recent advances in the field of multi-pump polariton superfluids \cite{Cristofolini}, we theoretically propose and experimentally realize a scheme to generate stable chains of vortices all having the  same sign in a coherently driven polariton superfluid. The scheme consists of a Laguerre-Gauss (LG) laser beam  with orbital angular momentum $\ell$ and an additional Gaussian beam (G) with zero angular momentum located at the center.

\begin{figure*}[ht!]
\includegraphics[scale=0.26]{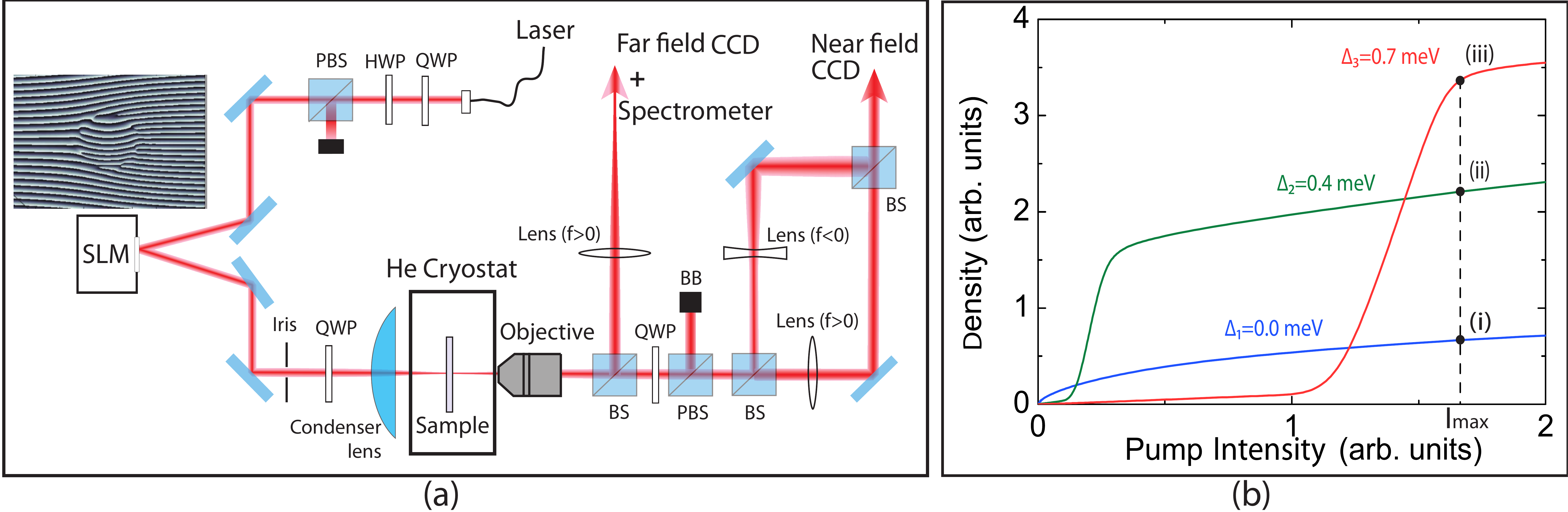}
\caption{(color online) \textbf{Setup and bistability} (a) - Scheme of the experimental setup. The pumps are prepared with a pure-phase SLM and sent on the cryostat containing the sample. The sample emission is then collected and treated for detection in real space, momentum space and energy. The inset shows the hologram used on the SLM. (b) - Scheme of the different regimes of bistability used (only the curve obtained when decreasing the density is shown). A stronger bistability (higher $\Delta$) provides higher densities for high pump intensities. $I_{max}$ indicates the pump peak intensity.}
\label{setup}
\end{figure*}

{\it Experimental setup.}$-$The laser used to resonantly excite the sample is a CW single mode Ti:Sa locked in frequency. Resonant pumping creates a low-density exciton gas, which allows us to neglect interactions between polaritons and the exciton reservoir, in contrast with experiments performed under non-resonant pumping \cite{Cristofolini}. The laser is then tuned to be quasi-resonant with the ground state of the lower polariton (LP) branch ($\sim \unit{837}{\nano\meter}$), such that the detuning with respect to the bottom of the LP branch is given by $\Delta=\omega_{laser}-\omega_{pol}$, where $\omega_{laser}$ and $\omega_{pol}$ are respectively the pump and non-renormalized polariton energies at normal incidence $k=\unit{0}{\micro\meter^{-1}}$. A single-mode fiber selects the $\textrm{TEM}_{00}$ mode. A series of quarter- and half-wave plates set the polarization to vertical and the remaining polarization fluctuations are cut by a polarizing beam splitter (PBS). The collimated laser then is sent on a Spatial Light Modulator (SLM), which allows us to arbitrarily modify the spatial phase profile of the beam. By sending a specifically designed phase hologram to the SLM, we can create beams with well-defined intensity and phase profiles. Making use of a hologram, we create a coherent superposition of a Laguerre-Gauss (LG) beam of orbital momentum $\ell=8$ and a Gaussian (G) beam of zero orbital momentum at the center (see Fig.\ref{setup} (a)). The relative size of the beams, intensities and focalizations are determined by the hologram. The relative sizes are chosen with the G beam smaller than the LG so that only their tails can overlap, making the interference pattern very weak. A grating hologram is added to the pump hologram to spatially deviate the first order reflection forming the pump, allowing us to block the light resulting from zero-order reflections. To avoid spin-dependent phenomena, the ($LG_{0}^{8}+LG_{0}^{0}$) polarization is set to circular with a quarter-wave plate before being focused on the sample by an aspherical condenser.

{\it Sample.}$-$The sample is a $2\lambda$-GaAs planar microcavity containing three GaAs-InGaAs quantum wells, resulting in a polariton Rabi splitting of $\unit{5.1}{\milli\electronvolt}$. The cavity finesse is $F=3000$, which results in a polariton lifetime of about $\tau\approx \unit{15}{\pico\second}$. The cavity is wedged in one direction, providing a large choice of cavity-exciton detunings by pumping on different positions on the sample. To enhance the polariton-polariton interactions, we use a cavity-exciton detuning of $\delta=\omega_C-\omega_X=\unit{1}{\milli\electronvolt}$ ($\omega_{X(C)}$ is the excitonic (cavity) energy at normal incidence, $k=\unit{0}{\micro\meter^{-1}}$), as this reinforces the exciton component of polaritons, therefore enhancing the nonlinear effects. The microcavity is cooled down to \unit{5}{\kelvin} in a cryostat and the measurements are taken in transmission. Above some critical value of $\Delta$, a bistable behavior appears \cite{Baas}. Increasing $\Delta$ and working on the upper bistability branch allows us to increase the polariton density. This is necessary to reach the regime where nonlinearities dominate. We use the bistability to control the density as follows: in the upper bistability branch, at constant pump power $I_p=\unit{300}{\milli\watt}$, we modify $\Delta$. We consider three different cases: low density ($\Delta_1\approx\unit{0}{\milli\electronvolt}$, point (i) in Fig.\ref{setup} (b)), high density far from the bistability threshold ($\Delta_2=\unit{0.4}{\milli\electronvolt}$, point (ii) in Fig.\ref{setup} (b)) and near the bistability threshold ($\Delta_3=\unit{0.7}{\milli\electronvolt}$, point (iii) in Fig.\ref{setup} (b)).

{\it Detection.}$-$The detection is made simultaneously in real space, momentum space and energy. An objective collects the sample emission. CCD cameras are used for direct imaging of the real space and the momentum space while the energy (wavelength) is measured with a spectrometer. We only collect circularly-polarized light, therefore filtering out any spin-flip effect. The polariton phase is measured with an off-axis interferometry setup: a beam splitter divides the real space image into two parts, one of which is expanded to generate a phase reference beam. The selection of the Gaussian part at the center of the image ensures a flat phase reference, which is used to make an off-axis interference pattern. With this method, the vortex position is independent from the phase of the reference beam \cite{bolda}. The actual phase map is then numerically reconstructed with a standard off-axis phase detection method. \par

\begin{figure}
\includegraphics[scale=0.26]{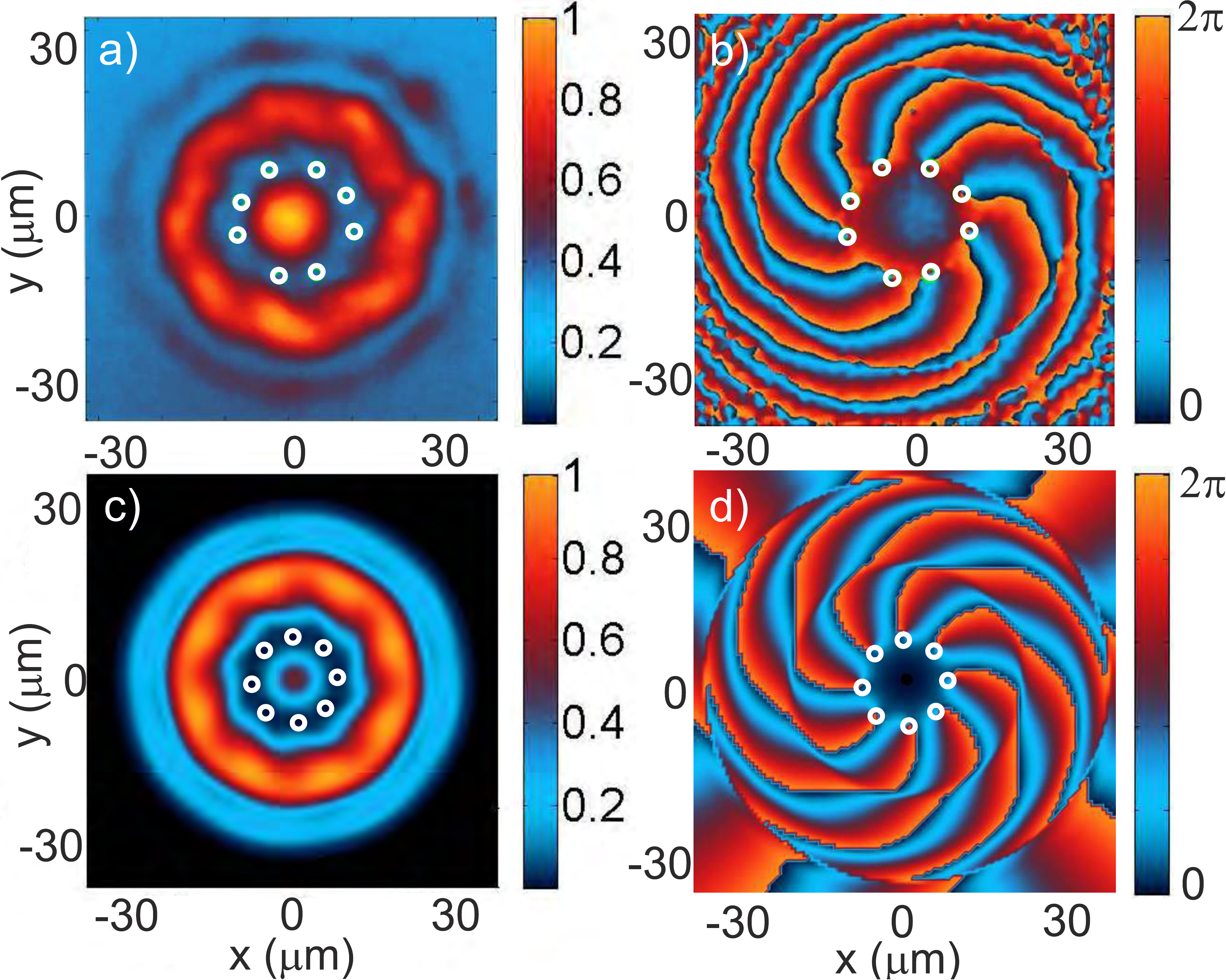}
\caption{(color online) \textbf{Linear regime} - Experimental (above) and theoretical (below) images of the polariton field in the linear regime (low densities, $\Delta\approx\unit{0}{\milli\electronvolt}$). The intensity $\vert\psi\vert^2$ (normalized by the peak value) is shown on the left panels while the phase $\arg(\psi)$ is shown on the right panels. Vortices are indicated by white circles both in the intensity and the phase. At low densities, the nonlinear interactions are negligible: the pattern is the result of optical interferences, fixing the vortex positions.}
\label{linear}
\end{figure}

{\it Theory.}$-$ In order to describe the configuration under study, we solve numerically the driven-dissipative scalar Gross-Pitaevskii equation, which in the parabolic approximation reads
\begin{equation}
i\hbar \frac{\partial \psi}{\partial t}=-\frac{\hbar^2\nabla^2}{2m}\psi-\frac{i\hbar}{2\tau}\psi+\alpha_1\vert \psi\vert^2\psi+P(\mathbf{r})e^{i \Delta t}.
\label{GP}
\end{equation}
Here, $P(\mathbf{r})=P_{LG}(\mathbf{r})+P_G(\mathbf{r})$ with $P_{LG} (\mathbf{r})=A_{1}(r/R_1)^{\ell/2} e^{-r^2/\sigma_1^2}e^{i\ell\varphi}$ and $P_{G}(\mathbf{r})=A_2 e^{-r^2/R_2^2}$. $A_1$ and $A_2$ are the amplitudes of the pumping lasers. To match the experimental configuration, we set $\sigma_1=5.0$ $\mu$m, $R_1=10.0$ $\mu$m and $R_2=3.0$ $\mu$m.

\begin{figure}
\includegraphics[scale=0.26]{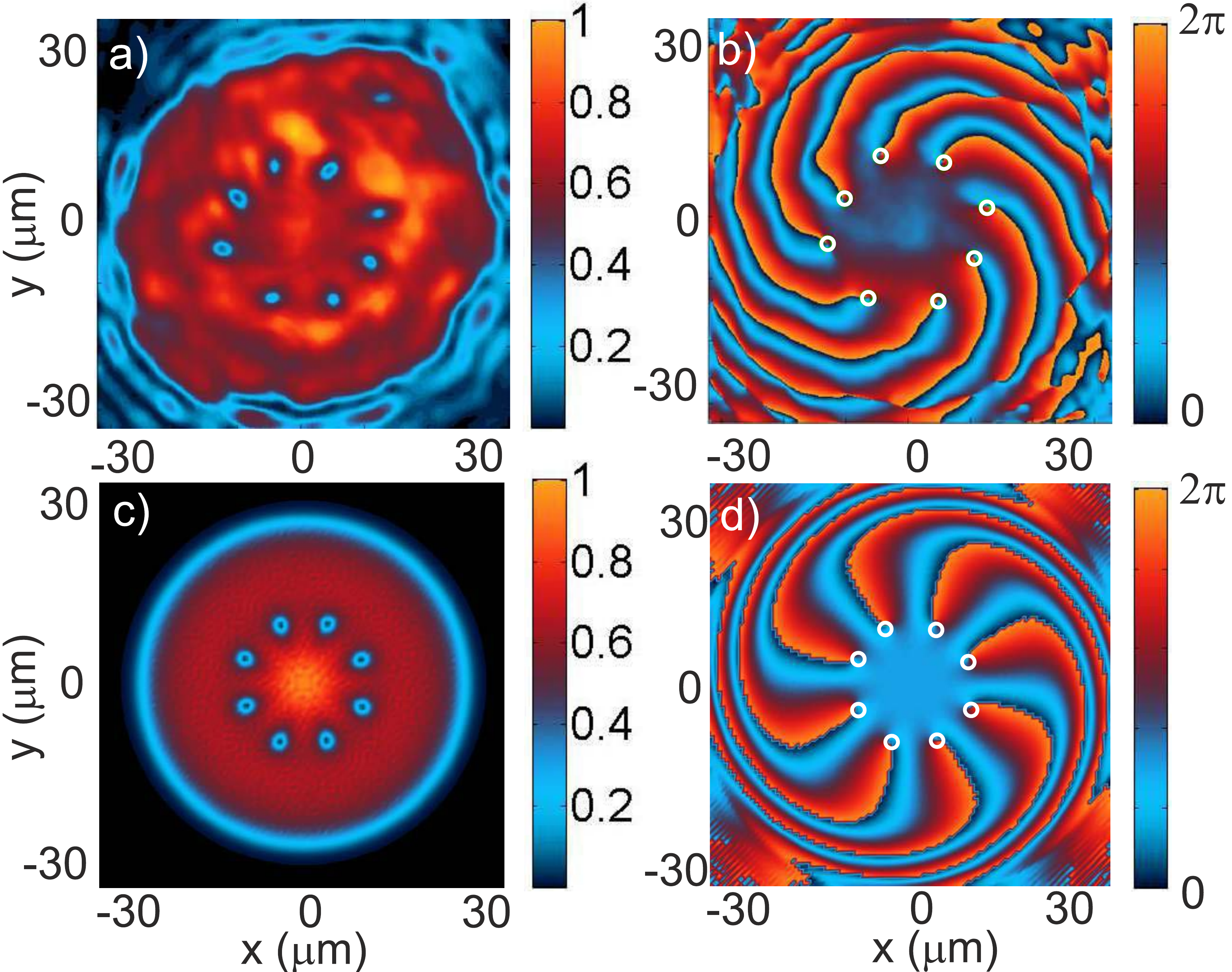}
\caption{(color online) \textbf{Nucleation of vortices and chain formation} - Experimental (above) and theoretical (below) images of the polariton field in the nonlinear regime (high densities, $\Delta=\unit{0.4}{\milli\electronvolt}$). The intensity $\vert\psi\vert^2$ (normalized by the peak value) is shown on the left panels while the phase $\arg(\psi)$ is shown on the right panels. Vortices are clearly visible in the intensity as eight zero-density dips in a) and c) and are indicated by white circles in the phase (diagrams b) and d)). The polariton density is high enough to enter the superfluid regime and to suppress the pump interference through nonlinear interactions.}
\label{nonlinear}
\end{figure}

{\it Results.}$-$In the linear regime ($\Delta=\Delta_1\approx\unit{0}{\milli\electronvolt}$), we observe both experimentally (Fig.\ref{linear} a),b)) and theoretically (Fig.\ref{linear} c),d)) a pattern resulting from the optical interference between the LG and G beams. This interference pattern consists in an eight-lobbed spiral (Fig.\ref{linear} a),c)) and its phase contains a ring with eight optical defects (Fig.\ref{linear} b),d)). The annular vortex chain visible in Fig.\ref{linear} is thus imposed by the pump phase. 
As the density increases, the nonlinear behavior of polaritons unveils, with a progressive deformation and disappearance of the interferences. For $\Delta=\Delta_2=\unit{0.4}{\milli\electronvolt}$ (point (ii) in Fig.\ref{setup} (b)), the interference pattern shown in Fig.\ref{nonlinear} is reduced to round-shaped dips of quasi-zero density containing the phase singularities: the elementary vortices carrying the injected angular momentum. We observe a ring of same-sign vortices in a coherent superfluid of polariton, quite differently from the spiral interference pattern present in the linear regime. The absence of interference pattern means that the polaritons phase is now different from the one of the pump and is modified through the nonlinear interactions, generating additional features compared to an optical interference pattern. This is in agreement with other resonant and non-resonant pumping experiments done in absence of angular momentum \cite{Cristofolini, PRB-Hivet}.

\begin{figure}[t!]
\includegraphics[scale=0.43]{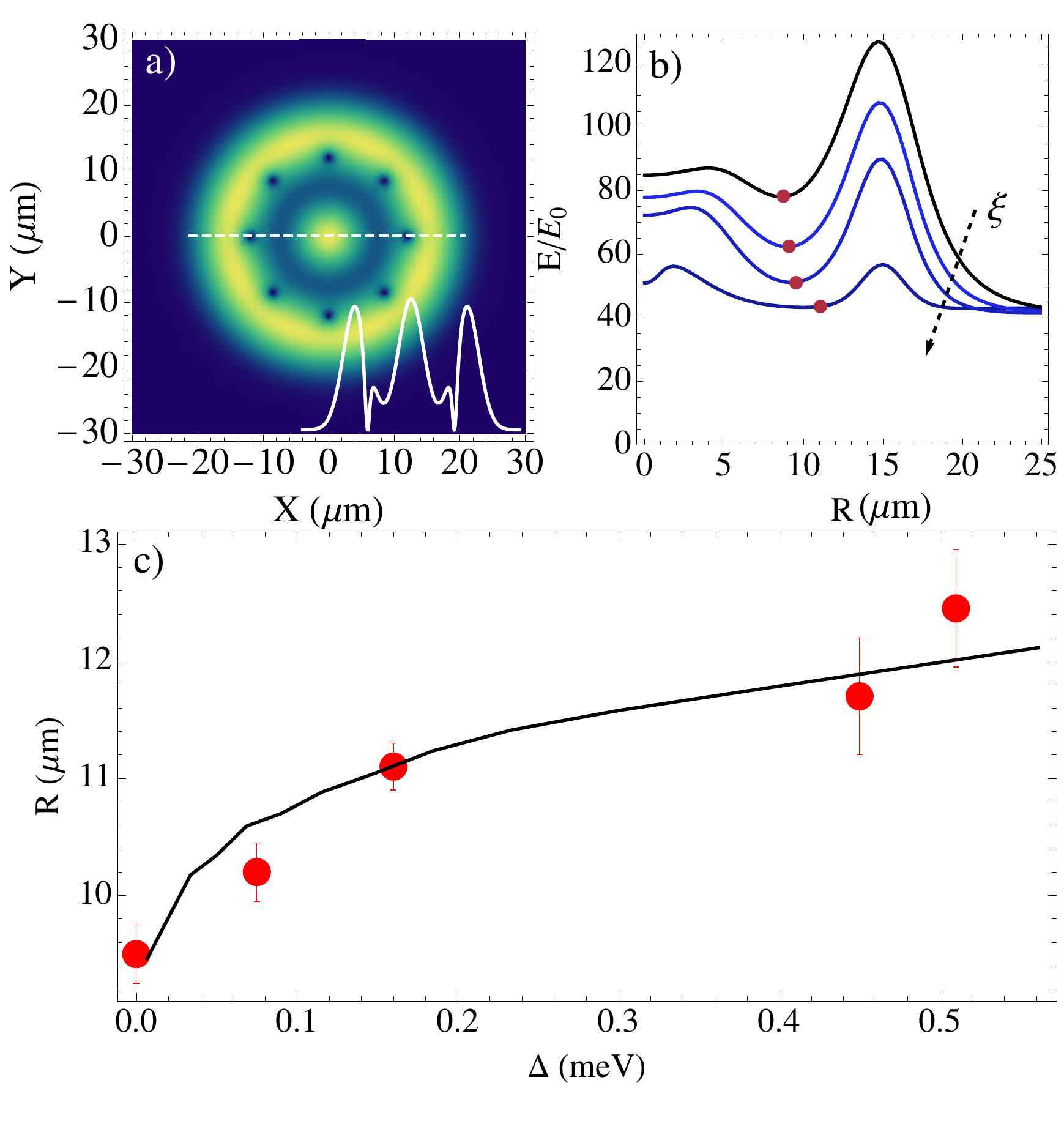}
\caption{(color online) \textbf{Vortex positions} (a) - Variational wave function in Eq. \ref{ansatz} used to determine the vortex chain radius $R$. The inset shows the radial profile for a cut at $Y=0$. (b) - Variational energy $E$ (in units of $E_0=\hbar^2\rho/(4\pi m)$) showing a minimum (red dots) for different values of $\xi$. From top to bottom: $\xi=(3.0, 2.0, 1.0, 0.5)$ $\mu$m. (c) - Average chain radius $R$ obtained as a function of the pump detuning $\Delta$ obtained from the variational method described in the text (solid line) and the corresponding experimental data (dots).}
\label{radius}
\end{figure}

It is important to remark that the spatial freedom of the vortices yields a quantitative test of our model. As mentioned in the introduction, nonlinear vortices exhibit a radial phase freedom: despite the fact that the azimuthal position of the vortices are locked by the pump phase, their radial position is modified as the density $\rho=\vert \psi\vert^2$ increases.  In order to model the ring radius $R$, we use a variational method. We describe the vortex chain solution with the variational ansatz 
\begin{equation}
\Psi[R]=\psi_\mathrm{TF}(r)\prod_{i=1}^8f_V(\mathbf{r}-\mathbf{r}_i),  \quad f_V(r_i,\theta_i)=\sqrt{\frac{r_i^2}{r_i^2+\xi^2}}e^{i\theta_i}, 
\label{ansatz}
\end{equation}
where $\psi_\mathrm{TF}(r)=\rho^{1/2}[P_1(r)/A_1+P_2(r)/A_2]$ is the Thomas-Fermi density profile induced by the pump, $\mathbf{r}_i=R(\cos\theta_i,\sin\theta_i)$ is the position of each vortex in the chain, and $\xi=\hbar/\sqrt{m\rho\alpha_1}$ is the average healing length. In Fig. \ref{radius} a), we plot the variational profile given by Eq. (\ref{ansatz}). The value of $R$ can then be extracted by minimizing the total energy $E[R]=E_\mathrm{kin}+E_\mathrm{int}$ and taking the physically relevant solution of the condition $\delta E/\delta R=0$, where
\begin{equation}
\begin{array}{c}
\displaystyle{E_\mathrm{kin}=\int d\mathbf{r} \left(\Psi^* \frac{\hbar^2\nabla^2}{2m}\Psi\right),} \\
\displaystyle{E_\mathrm{int}=\int d\mathbf{r}\left[\left(\vert\Psi\vert-\rho\right)^2\vert\Psi\vert^2 \right]}.
\end{array}
\end{equation}
In Fig. \ref{radius} b), we plot the energy $E$ as a function of $R$ for different values of $\xi~\sim \rho^{-1/2}$. We observe that $R$ decreases (increases) as the value of $\xi$ ($\rho$) is increased. Since $\rho$ increases with $\Delta$, $\xi$ decreases with $\Delta$; it follows that $R$ increases when we increase $\Delta$. Fig.\ref{radius} c) shows $R$ as a function of $\Delta$ calculated by this method and the comparison with experimental data. As expected, the higher the density, the further from the center the vortices migrate. This behavior and the agreement between the variational method and experiment are a clear indication of the phase freedom obtained when interactions dominate, a feature that is independent from the optical interference.

\begin{figure}[b!]
\includegraphics[scale=0.37]{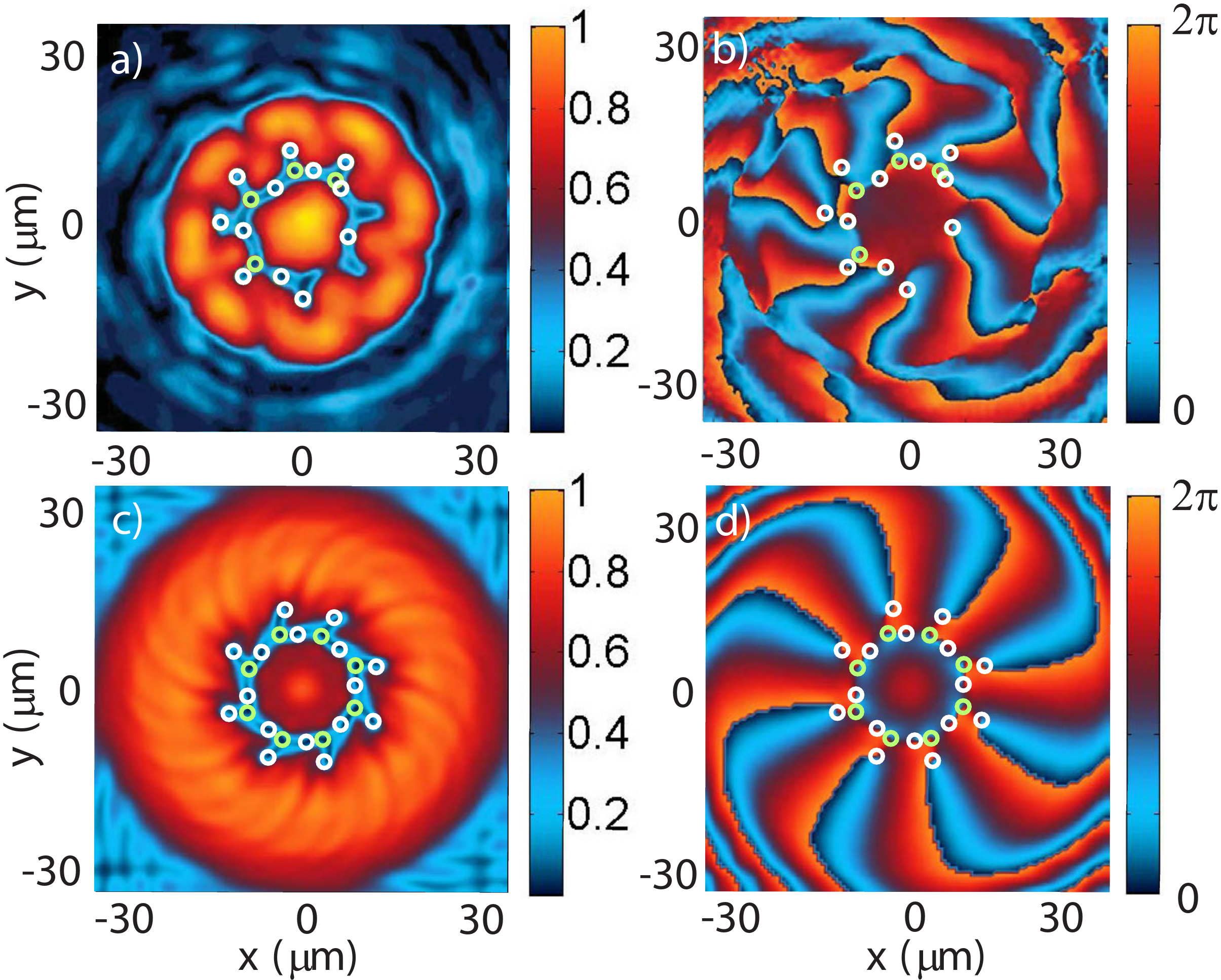}
\caption{(color online) \textbf{Phase instability} - Experimental (above) and theoretical (below) images of the polariton field when the bistability threshold is close to the pump maxima (very high densities, $\Delta=\unit{0.7}{\milli\electronvolt}$). The intensity $\vert\psi\vert^2$ (normalized by the peak value) is shown on the left panels while the phase $\arg(\psi)$ is shown on the right panels. Vortices are indicated by white circles and antivortices by green circles both in the intensity and the phase. Here only the maxima of the pump are on the upper bistability branch (resonant pumping) and produce a significant polariton density. The nonresonant zone between the LG and G pumps is wide enough for the polariton fluid to hydrodynamically nucleate into vortex-antivortex pairs, while preserving the annular vortex chain.}
\label{sun}
\end{figure}

In the upper bistability branch, the polariton energy is renormalized through self-interaction to the pump energy, so that the pumping is resonant and yields high polariton densities. This is not the case in the lower bistability branch,  where the pumping is non-resonant and inefficient. For large $\Delta$ the low pump intensity regions, on the lower bistability branch, are far from resonance as can be seen in Fig.\ref{setup} (b). In these regions, non-resonant pumping yields a negligible polariton population and can be considered as a pump-free region. For exemple near the threshold, for $\Delta=\Delta_3=\unit{0.7}{\milli\electronvolt}$, point (iii) in Fig.\ref{setup} (b)), a large area between the LG and G pumps is not pumped. Although no polariton is created in this area, the density is not zero due to polaritons propagating from the pumped to the non-pumper area. The phase in this region is free to evolve, which explains the radius chain expansion previously mentioned. Moreover when the density in this pump-free area is large enough and when this area is at least of the size of the vortex core $\sim \xi$, we observe the spontaneous nucleation of vortex-antivortex pairs. Four pairs are experimentally visible in Fig.\ref{sun} and eight in the theoretical figure. They form a quasi-continuous, low-density ring inside the vortex chain described above. This is due to a hydrodynamic instability of the same nature as in the $\ell=0$ case \cite{manni}, but here each pair formation is stimulated by the presence of a vortex which acts as a defect. This feature further proves that the vortex distribution is not due to optical interference in the superfluid regime, but rather that it evolves with the density. \par

{\it Conclusion.}$-$In the present work, we resonantly inject polaritons with a given total angular momentum and observe the formation of a ring of quantized single-charged vortices. For the first time, a regular ring pattern of elementary vortices of the same sign is reported in a polariton superfluid. In the superfluid regime, the radial position is not determined by the pump but rather depends on the polariton density. Experimental and theoretical indications of this property, due to very strong nonlinear interactions, are provided through the system hydrodynamical behavior \cite{comment}.
The mechanism leading to the creation of vortex chains results from combination of the saturation of the radial counterflow instability with the injection of angular momentum in a limited region of the space. We expect that the present scheme will pave the stage to study a series of new vortex collective phenomena that have not been possible so far. One important example are the Tkachenko modes \cite{fetter, sonin, coddington} in driven-dissipative systems. Also, we may be interested in studying quantum features of one-dimensional properties of vortex chains. Such a system may be investigate in the context of long-range interacting Luttinger liquids \cite{schulz, tercaswig}.

{\it Acknowledgments.} We acknowledge the financial support of the ANR Quandyde (ANR-11-BS10-001), ANR Labex GANEX (ANR-11-LABX-0014) and IRSES POLAPHEN (246912).

\FloatBarrier


\begin{thebibliography}{30}
\bibitem{comment}
A characteristic sign of this phase constrain disappearance is the vanishing of the inference pattern in the superfluid regime, as previously reported \cite{PRB-Hivet, Cristofolini}.
\end{thebibliography}
\end{document}